\documentclass{article}
\usepackage{spconf,amsmath,amssymb}
\usepackage{graphicx}
\usepackage{bm,cite,mathtools}
\usepackage{subfig}
\newcommand{\e}{\bm{e}}
\renewcommand{\d}{\bm{d}}
\renewcommand{\r}{\bm{r}}
\newcommand{\x}{\bm{x}}
\newcommand{\y}{\bm{y}}
\newcommand{\G}{\bm{G}}
\newcommand{\W}{\bm{W}}
\newcommand{\A}{\bm{A}}
\newcommand{\K}{\bm{K}}
\newcommand{\bxi}{\bm{\xi}}
\renewcommand{\H}{\mathsf{H}}
\newcommand{\Trans}{\mathsf{T}}
\newcommand{\ue}{u_{\mathrm{e}}}
\newcommand{\up}{u_{\mathrm{p}}}
\newcommand{\us}{u_{\mathrm{s}}}
\newcommand{\hue}{\hat{u}_{\mathrm{e}}}
\newcommand{\hup}{\hat{u}_{\mathrm{p}}}
\newcommand{\hus}{\hat{u}_{\mathrm{s}}}
\newcommand{\bmeta}{\bm{\eta}}
\newcommand{\bmze}{\bm{z}_{e}}

\newcommand{\bmzd}{\bm{z}_{d}}
\title{Spatial active noise control based on individual kernel interpolation of primary and secondary sound fields}
\name{Kazuyuki Arikawa, Shoichi Koyama, and Hiroshi Saruwatari}
\address{The University of Tokyo, 7-3-1 Hongo, Bunkyo-ku, Tokyo 113-8656, Japan}
\begin{document}
\ninept
\maketitle
\begin{abstract}
A spatial active noise control (ANC) method based on the individual kernel interpolation of primary and secondary sound fields is proposed. Spatial ANC is aimed at cancelling unwanted primary noise within a continuous region by using multiple secondary sources and microphones. A method based on the kernel interpolation of a sound field makes it possible to attenuate noise over the target region with flexible array geometry. Furthermore, by using the kernel function with directional weighting, prior information on primary noise source directions can be taken into consideration. However, whereas the sound field to be interpolated is a superposition of primary and secondary sound fields, the directional weight for the primary noise source was applied to the total sound field in previous work; therefore, the performance improvement was limited. We propose a method of individually interpolating the primary and secondary sound fields and formulate a normalized least-mean-square algorithm based on this interpolation method. Experimental results indicate that the proposed method outperforms the method based on total kernel interpolation. 

\end{abstract}
\section{Introduction} 
\label{sec:intro}

Active noise control (ANC) is a technique of canceling unwanted noise by emitting anti-noise signals from loudspeakers~\cite{Nelson:ACS,kuo1999active,kajikawa2012recent}. In general, adaptive filtering algorithms are applied to obtain driving signals of secondary loudspeakers for attenuating the primary noise by using error and reference microphone signals. When applying this conventional ANC technique in a three-dimensional (3D) space, the primary noise at error microphone positions can be reduced, but there is no guarantee that the primary noise is reduced in the region between these positions. 

In recent studies, spatial ANC methods, which aim to reduce incoming noise in a spatial target region, have been intensively investigated~\cite{Zhang:ANC, zhangg2016sparse,Bu:ACM2018,maeno2019spherical,sun2019active,Sun:ICASSP2019,Ito:ICASSP2019,Koyama:IEEE_ACM_J_ASLP2021} owing to recent advances in sound field analysis and synthesis methods~\cite{poletti2005three,spors2008theory,wu2008theory,Koyama:IEEE_J_ASLP2013,Ueno:IEEE_SPL2018,Ueno:IEEE_J_SP_2021}. Since most spatial ANC methods overly depend on the spherical/cylindrical harmonic analysis of a sound field~\cite{Zhang:ANC, Bu:ACM2018, maeno2019spherical, sun2019active, Sun:ICASSP2019}, applicable array configurations of microphones are limited. On the other hand, the spatial ANC method based on the kernel interpolation of a sound field has flexibility in array configurations~\cite{Ito:ICASSP2019,Koyama:IEEE_ACM_J_ASLP2021}. Furthermore, the prior identification stage required in virtual sensing techniques~\cite{Moreau:2008,liu2009virtual} is unnecessary. 

In kernel-interpolation-based spatial ANC, it is essential to appropriately design a kernel function for estimating a continuous sound field from a discrete set of microphones. In \cite{Ito:ICASSP2020}, a kernel function with directional weighting was proposed to incorporate prior information on primary noise source directions. The estimation accuracy can be enhanced for predefined directions by appropriately setting parameters of the weight thus leading to a larger regional noise reduction than the method using a uniform weight. 

An issue in this interpolation procedure is that the directional weight for the primary noise source is applied to the total sound field. Since the total sound field is essentially a superposition of primary and secondary sound fields in spatial ANC, simply weighting only in the direction of the primary noise source is inappropriate in most cases. Therefore, we propose a method of individually interpolating the primary and secondary sound fields. By separately handling the primary and secondary sound fields with the directional weight for each primary/secondary source, the interpolation accuracy can be further improved from the previous method based on a single directional weight for the total sound field.  We also formulate a normalized least-mean-square (NLMS) algorithm for feedforward spatial ANC based on this individual interpolation. Numerical experiments are conducted to validate the effectiveness of the proposed method. Although all the formulations are derived in the frequency domain for simplicity, time-domain algorithms can be readily obtained in a manner similar to that in our prior works~\cite{Brunnstrom:ICASSP2021, Koyama:IEEE_ACM_J_ASLP2021}.

\section{Problem statement and prior works}

\subsection{Problem statement}
\label{sec:conventional}

\begin{figure}[tb]
  \centering
  \centerline{\includegraphics[width=0.9\columnwidth]{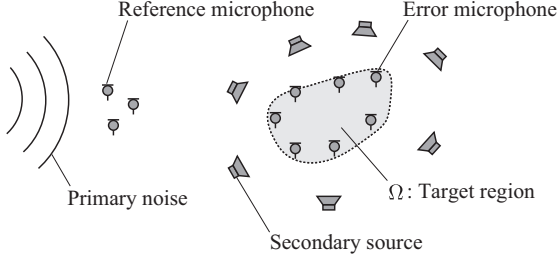}}
\caption{Schematic diagram of feedforward spatial ANC.}
\label{fig:2-01}
\end{figure}

Suppose that a target region $\Omega \subset \mathbb{R}^3$ is set and $M$ error microphones are placed in $\Omega$. $L$ secondary sources (loudspeakers) and $R$ reference microphones are placed in the exterior region of $\Omega$. The goal of spatial ANC is to reduce incoming noise from primary noise sources over $\Omega$ by generating an anti-noise field using the secondary sources with measurements of the error and reference microphones. A schematic diagram of this feedforward spatial ANC framework is shown in Fig.~\ref{fig:2-01}.


We denote the frequency-domain driving signals of the secondary sources and the observed signals of the reference and error microphones at angular frequency $\omega$ as $\y(\omega)\in \mathbb{C}^{L}$, $\x(\omega) \in \mathbb{C}^{R}$, and $\e(\omega)\in \mathbb{C}^{M}$, respectively. Hereafter, the argument $\omega$ is omitted for notational simplicity. When denoting the primary noise at the error microphone positions as $\d(\omega) \in \mathbb{C}^{M}$, the error microphone signals are represented as
\begin{align}
    \e = \d +\G \y, 
    \label{eq:e_compose}
\end{align}
where  $\G \in \mathbb{C}^{M \times L}$ consists of the transfer functions from the secondary sources to the error microphones. We assume that the estimate of $\G$, $\hat{\G}$, is given by measuring them in advance. It will also be possible to incorporate online secondary path modeling~\cite{Chan:IEEE2012}. The driving signals $\y$ are obtained as
\begin{align}
    \y = \W \x,
    \label{eq:y}
\end{align}
where $\W \in \mathbb{C}^{L \times R}$ is the control filter matrix, which is adaptively changed by adaptive filtering algorithms using $\e$.


\subsection{Spatial ANC based on kernel interpolation of sound field}

In \cite{Ito:ICASSP2019, Ito:ICASSP2020}, the cost function for spatial ANC is defined as the acoustic potential energy inside $\Omega$ as
\begin{align}
    J = \int_{\Omega} |\ue(\r)|^2\,\mathrm{d}\r, 
    \label{eq:J}
\end{align}
where $\ue \colon \mathbb{R}^3 \to \mathbb{C}$ is the continuous pressure distribution inside $\Omega$ during the adaptive process. To estimate $\ue$ from a discrete set of measurements $\e$, the kernel interpolation method for a sound field~\cite{Ueno:IEEE_SPL2018, Ueno:IEEE_J_SP_2021} is applied. The pressure field at arbitrary position $\r\in\Omega$ is estimated as
\begin{align}
    \hue(\r) &= \bm{z}_{e}(\bm{r})^{\Trans} \e
    \label{eq:hue_trad}
\end{align}
with the interpolation filter 
\begin{align}
    \bm{z}_{e}(\bm{r}) & \coloneqq \left[ (\K + \lambda \bm{I}_M)^{-1} \right]^{\Trans}\bm{\kappa}(\bm{r}),
  \label{eq:interp_filter}
\end{align}
where $(\cdot)^{\Trans}$ denotes the transpose, $\lambda$ is a positive regularization parameter, and the matrix $\K \in \mathbb{C}^{M \times M}$ and vector $\bm{\kappa}(\r) \in \mathbb{C}^M$ consist of the kernel function $\kappa(\cdot, \cdot)$. Specifically, the $(m, m^{\prime})$th element of $\K$ (i.e., the Gram matrix) is $\kappa(\r_m ,\r_{m^{\prime}})$ and the $m$th element of $\bm{\kappa}(\r)$ is $\kappa(\r ,\r_m)$ with the position of the $m$th error microphone $\r_m$. By substituting \eqref{eq:hue_trad} into \eqref{eq:J}, the cost function is represented by a quadratic form of $\e$ as
\begin{align}
    J &= \e^{\H} \A \e,
    \label{eq:J_int}
\end{align}
where $(\cdot)^{\H}$ denotes the conjugate transpose and $\A$ is the Hermitian interpolation matrix defined as
\begin{align}
    \A &= \int_{\Omega} \bm{z}_{e}^{\ast}(\bm{r})\bm{z}_{e}^{\Trans}(\bm{r})\,\mathrm{d}\bm{r} \notag\\
    &= \bm{P}^{\H} \left[ \int_{\Omega} \bm{\kappa}(\r)^{\ast} \bm{\kappa}(\r)^{\Trans} \mathrm{d}\r  \right] \bm{P}.
    \label{eq:A}
\end{align}
Here, $(\cdot)^*$ denotes the complex conjugate and $\bm{P}\coloneqq (\K + \lambda \bm{I}_M)^{-1}$. The integral in \eqref{eq:A} is usually computed by numerical integration with some exceptions~\cite{Ito:ICASSP2019,Ito:ICA2019}. Thus, the NLMS algorithm for updating $\W$ to minimize $J$ is derived by using the gradient $\partial J/\partial \W^{\ast} = \hat{\G}^{\H}\bm{A}\e\x^{\H}$ as
\begin{align}
    \W&(n+1) \notag \\
  &= \W(n) - \frac{\mu_0}{\|\hat{\G}^{\H} \A \hat{\G}\|_2 \|\x(n)\|_2^2 + \epsilon}{\hat{\G}}^{\H} \A \e(n) \x(n)^{\H}, \label{eq:trad_grad}
\end{align}
where $\|\cdot\|_2$ for matrices denotes the maximum singular value, $n$ denotes the time index, $\mu_0 \in (0, 2)$ is a normalized step size parameter, and $\epsilon > 0$ is a regularization parameter to avoid zero division.
\section{Spatial ANC based on individual kernel interpolation}

\subsection{Kernel function with directional weighting and its limitation}

The choice of the kernel function $\kappa(\cdot, \cdot)$ in \eqref{eq:interp_filter} is particularly important because the estimation accuracy of $\ue$ is highly dependent on it. Ito~et al.~\cite{Ito:ICASSP2020} proposed a kernel function with directional weighting to incorporate prior knowledge on primary noise source directions into the estimation. We here assume that a single primary noise source direction is given, which is denoted by $\bm{\eta}$. The kernel function is defined as the weighted integral of plane waves over the unit sphere $\mathbb{S}_2$ as
\begin{align}
    \kappa(\r_1, \r_2) = \frac{1}{4\pi}\int_{\mathbb{S}_2}\gamma(\bxi) \mathrm{e}^{\mathrm{j}k\bm{\bxi}^{\Trans}(\r_1 - \r_2)}\,\mathrm{d}\bxi, \label{eq:kappa}
\end{align}
where $k\coloneqq \omega/c$ is the wave number defined with the sound speed $c$, and the directional weighting function $\gamma \colon \mathbb{S}_2 \to \mathbb{R}$ originating from the von Mises--Fisher distribution~\cite{mardia2009directional} is defined as
\begin{align}
    \gamma(\bxi) = \mathrm{e}^{\beta \bxi^{\Trans}\bm{\eta}}. \label{eq:gamma}
\end{align}
Here, $\beta \geq 0$ is a parameter used to control the sharpness of the weights in the direction of $\bmeta$. Then, the kernel function $\kappa$ is derived as
\begin{align}
    \kappa(\r_1, \r_2) = j_0 \left( \sqrt{(\mathrm{j}\beta \bm{\eta} - k\r_{12})^{\Trans}(\mathrm{j}\beta \bm{\eta} - k\r_{12})} \right),
    \label{eq:ker_dir}
\end{align}
where $\r_{12} \coloneqq \r_2 - \r_1$.
Note that the sound field interpolated by using this kernel function satisfies the Helmholtz equation. It was shown in \cite{Ito:ICASSP2020} that by appropriately setting $\bm{\eta}$ and $\beta$, a large noise power reduction is achieved  compared with the case of a uniform weight $\gamma(\bm{\xi})=1$~\cite{Ito:ICASSP2019}; however, the performance improvement was not significant.


The main issue of the above interpolation procedure is that the directional weighting for the primary noise source is applied to the total sound field $\ue$. The sound field $\ue$ is essentially a superposition of the primary noise field $\up$ and the secondary sound field generated by the secondary sources $\us$. Nevertheless, both $\up$ and $\us$ are interpolated in total as 
\begin{align}
    \hue(\r) &= \hup(\r) + \hus(\r) \notag \\
    &=\bmze(\r)^{\Trans}\d + \bmze(\r)^{\Trans}\G\y 
    \label{eq:hue_drawback}
\end{align}
with the kernel function using a single directional weighting in $\bmze(\r)$. However, the secondary sources generating $\us$ are not usually in the same direction as the primary noise source (see Fig.~\ref{fig:2-01}). This directional mismatch can limit the performance improvement of the method proposed in \cite{Ito:ICASSP2020}. We hereafter call the interpolation method based on \eqref{eq:hue_drawback} \textit{total kernel interpolation}.


\subsection{Individual kernel interpolation of primary and secondary sound fields}

We propose a sound field interpolation method for spatial ANC called \textit{individual kernel interpolation}, which is based on the individual interpolation of the primary and secondary fields. 

First, we decompose the error microphone signals $\e$ into the primary noise component $\hat{\d} = \e - \hat{\G} \y$ and secondary source component $\bm{s} = \hat{\G} \y$. Since $\hat{\G}$ is assumed to be given in advance and $\y$ is also a known value obtained using \eqref{eq:y}, $\hat{\d}$ and $\bm{s}$ can be obtained only from $\e$.



Next, we apply kernel interpolation to individually estimate $\up$ and $\us$. The primary field $\up$ is estimated from $\hat{\d}$ as
\begin{align}
    \hup(\r) = \bmzd(\r)^{\Trans}\hat{\d}, 
    \label{eq:up_prop}
\end{align}
where $\bmzd(\cdot)$ is obtained using \eqref{eq:interp_filter} with the directional weighting function designed for the primary noise source direction $\bmeta$. On the other hand, the secondary sound field $\us$ is estimated for each secondary source as 
\begin{align}
    \hus(\r) &= \sum_{l=1}^{L} \bm{z}_{y, l}(\r)^{\Trans} \hat{\G}_l y_l \notag\\
    &= \bm{\zeta}_y(\r)^{\Trans} \y, 
  \label{eq:us_prop}
\end{align}
where $\bm{\zeta}_y(\r) \coloneqq [\hat{\G}_1^{\Trans} \bm{z}_{y, 1}(\r), \ldots, \hat{\G}_L^{\Trans} \bm{z}_{y, L}(\r)]^{\Trans}$ with the $l$th column of $\hat{\G}$, $\hat{\G}_l$ ($l\in\{1,\ldots L\}$), and $\bm{z}_{y, l}(\cdot)$ is the interpolation filter given by \eqref{eq:interp_filter} with the direction of the $l$th secondary source $\bmeta_l$ used to compute the kernel function \eqref{eq:ker_dir}. Note that each $\bmeta_l$ is accurately known in typical spatial ANC applications. 

Finally, the total sound field $\ue$ is estimated using \eqref{eq:up_prop} and \eqref{eq:us_prop} as 
\begin{align}
    \hue(\r) &= \hup(\r) + \hus(\r) \notag \\
    &= \bmzd(\r)^{\Trans}\hat{\d} + \bm{\zeta}_y(\r)^{\Trans} \y. \label{eq:ue_prop}
\end{align}

\subsection{NLMS algorithm based on individual kernel interpolation}

\begin{figure}
    \centering
    \subfloat[Total-kernel-interpolation-based NLMS algorithm \label{fig:3-01a} ]{
    \includegraphics[width=0.67\columnwidth]{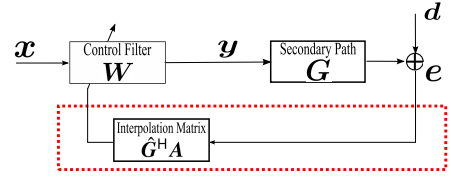}} \\
    \subfloat[Individual-kernel-interpolation-based NLMS algorithm \label{fig:3-01b} ]{
    \includegraphics[width=0.67\columnwidth]{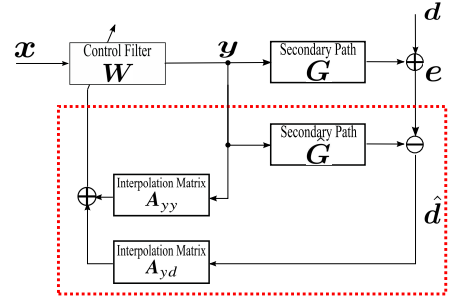}}
    \caption{Comparison of block diagrams of NLMS algorithms based on total and individual kernel interpolations. Red dotted rectangles indicate the blocks differentiating the two methods.} 
\label{fig:3-01}
\end{figure}

By substituting \eqref{eq:up_prop}, \eqref{eq:us_prop}, and \eqref{eq:ue_prop} into \eqref{eq:J}, the cost function $J$ is represented as
\begin{align}
    J = \hat{\d}^{\H} \A_{dd} \hat{\d} + {\y}^{\H} \A_{yd} \hat{\d} + \hat{\d}^{\H} \A_{yd}^{\H} \y + \y^{\H} \A_{yy} \y,
\end{align}
where $\A_{dd}\in\mathbb{C}^{M \times M}$, $\A_{yd}\in\mathbb{C}^{L \times M}$, and $\A_{yy}\in\mathbb{C}^{L \times L}$ are respectively defined as
\begin{align}
    \A_{dd} &\coloneqq \int_{\Omega} \bm{z}_{d}^{\ast}(\bm{r})\bm{z}_{d}^{\Trans}(\bm{r})\,\mathrm{d}\bm{r}\\       
    \A_{yd} &\coloneqq \int_{\Omega} \bm{\zeta}_{y}^{\ast}(\bm{r})\bm{z}_{d}^{\Trans}(\bm{r})\,\mathrm{d}\bm{r} \label{eq:A_yd}\\
    \A_{yy} &\coloneqq \int_{\Omega} \bm{\zeta}_{y}^{\ast}(\bm{r})\bm{\zeta}_{y}^{\Trans}(\bm{r})\,\mathrm{d}\bm{r}. \label{eq:A_yy}
\end{align}
An NLMS algorithm for individual-kernel-interpolation-based spatial ANC is derived using the gradient $\partial J/\partial \bm{W}^{\ast}$ as
\begin{align}
    \W(n+1) &= \W(n) - \frac{\mu_0}{\|\A_{yy}\|_2 \|\x(n)\|_2^2 + \epsilon}[\A_{yd} \e(n)\notag \\ 
  & \hspace{36pt} + (\A_{yy} - \A_{yd} \hat{\G})\W(n)\x(n)]\x(n)^{\H}. 
  \label{eq:prop}
\end{align}
The matrices $\A_{yd}$ and $\A_{yy}$ respectively defined in \eqref{eq:A_yd} and \eqref{eq:A_yy} can be represented in the same manner as the matrix $\bm{A}$ in \eqref{eq:A} and computed by numerical integration. Although the computational time for calculating $\A_{yd}$ and $\A_{yy}$ is greater than that for calculating $\A$ in the total-kernel-interpolation-based method, these matrices can be computed offline before the adaptive process starts. 
Figs.~\ref{fig:3-01}(a) and (b) show block diagrams of the NLMS algorithms based on the total and individual kernel interpolations, respectively, where red dotted rectangles indicate the blocks differentiating the two methods.

We also comment on the relationship between the total- and individual-kernel-interpolation-based NLMS algorithms. When the kernel functions for \eqref{eq:up_prop} and \eqref{eq:us_prop} in the individual kernel interpolation are the same with the same parameters $\bmeta$ and $\beta$, $\A_{yd}$ and $\A_{yy}$ are represented as $\A_{yd} = \hat{\G}^{\H} \A$ and $\A_{yy} = \hat{\G}^{\H} \A \hat{\G}$, respectively. Thus, the proposed method becomes identical to the method in \cite{Ito:ICASSP2020}. Therefore, individual-kernel-interpolation-based spatial ANC is a generalization of total-kernel-interpolation-based spatial ANC. 


\section{Experiments}

Experimental evaluations were performed by 3D free-field simulation to compare the proposed individual-kernel-interpolation-based spatial ANC (\textbf{Proposed}) with total-kernel-interpolation-based spatial ANC (\textbf{Total-KI-ANC}). We also evaluated the conventional multipoint pressure control (\textbf{MPC}), whose cost function is defined as the power of error microphone signals $\|\e\|_2^2$. The NLMS algorithm for \textbf{MPC} is obtained by replacing $\A$ in \eqref{eq:trad_grad} with $\bm{I}_M$. The parameter $\beta$ in \eqref{eq:ker_dir} was set to $10.0$ for the proposed method. In \textbf{Total-KI-ANC}, two values of $\beta$, $0.0$ and $2.0$, were investigated. $\beta=0.0$ corresponds to the uniform weight without the use of directional information, and $\beta=2.0$ was chosen from $(0.0, 10.0)$ by a grid search so that the regional noise power reduction was maximized. The integrals in \eqref{eq:A}, \eqref{eq:A_yd}, and \eqref{eq:A_yy} were computed by naive Monte Carlo integration with 2500 samples. The regularization parameter $\lambda$ in \eqref{eq:hue_trad} was set to $10^{-3}$. The parameters in the NLMS algorithm, $\mu_0$ and $\epsilon$ in \eqref{eq:trad_grad} and \eqref{eq:prop}, were set to $\mu_0 = 0.5$ and $\epsilon = 10^{-3}$, respectively. The sound velocity $c$ was $343$ $ \mathrm{m/s}$. 

\begin{figure}[t]
  \centering
  \centerline{\includegraphics[width=0.84\columnwidth]{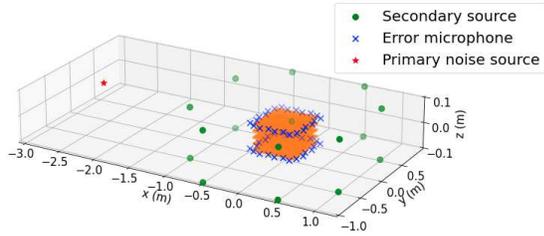}}
\caption{Experimental setup. The cuboid in orange indicates the target region.}
\label{fig:4-00}
\end{figure}

The target region $\Omega$ was a cuboid of $0.6~\mathrm{m} \times 0.6~\mathrm{m} \times 0.1~\mathrm{m}$, whose center was set at the coordinate origin as shown in Fig.~\ref{fig:4-00}. The numbers of secondary sources and error microphones were set to $L=16$ and $M=48$, respectively. The secondary sources were point sources regularly arranged on the borders of two squares of $2.0~\mathrm{m} \times 2.0~\mathrm{m}$ dimensions at the heights of $z=\pm 0.1~\mathrm{m}$. The error microphones were regularly placed on the borders of the top and bottom squares of $\Omega$, where every second microphone was shifted outwards by $0.03~\mathrm{m}$ to alleviate the forbidden frequency problem~\cite{Koyama:IEEE_ACM_J_ASLP2020}. A single point source was placed as the primary noise source at $(-2.8~\mathrm{m}, 0.3~\mathrm{m}, 0.0~\mathrm{m})$, and the reference microphone signal was directly obtained from the primary noise source. Gaussian noise was also added to the error microphone signal at each iteration so that the signal-to-noise ratio became $40~\mathrm{dB}$.

As an evaluation measure, we define the regional noise power reduction inside $\Omega$ as
\begin{align}
    P_{\mathrm{red}}(n) = 10\log_{10}\frac{\sum_j |\ue^{(n)}(\r_j)|^2}{\sum_j|\up^{(n)}(\r_j)|^2},
\end{align}
where $\r_j$ is the $j$th evaluation point in $\Omega$, and $\ue^{(n)}$ and $\up^{(n)}$ represent the total pressure field and primary noise field at the $n$th iteration, respectively. We set 1445 evaluation points inside $\Omega$.

\begin{figure}[t]
  \centering
  \centerline{\includegraphics[width=0.9\columnwidth]{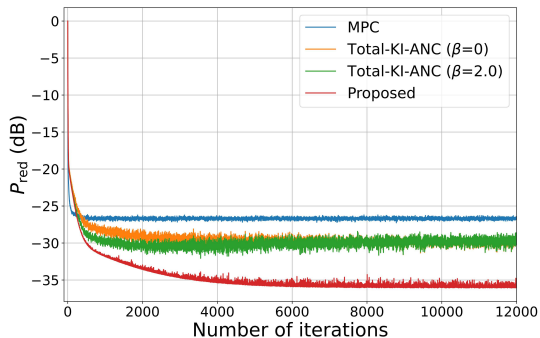}}
\caption{Regional noise power reduction at each iteration at $200~\mathrm{Hz}$.}
\label{fig:4-01}
\end{figure}

\begin{figure}[t]
  \centering
  \centerline{\includegraphics[width=0.9\columnwidth]{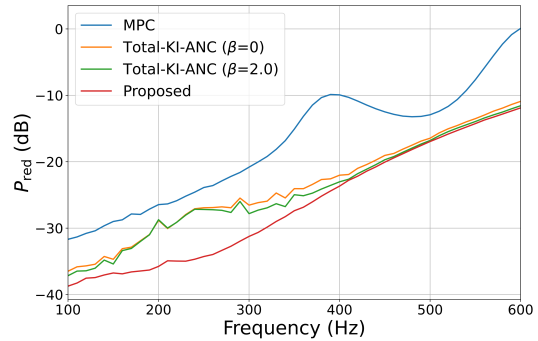}}
\caption{Regional noise power reduction after 12000 iterations with respect to frequency.}
\label{fig:4-02}
\end{figure}

We first show the result when the prior direction $\bmeta$ is exactly the same as the primary noise source direction. Fig.~\ref{fig:4-01} shows $P_{\textup{red}}$ at each iteration at 200 Hz. The difference in noise reduction between \textbf{Total-KI-ANC} for $\beta=0.0$ and $\beta=2.0$ was small. Among the four methods, the largest noise reduction was achieved by the proposed method. In Fig.~\ref{fig:4-02}, $P_{\mathrm{red}}$ at $n=12000$ is plotted with respect to the frequency from $100$ to $600~\mathrm{Hz}$ at intervals of $10~\mathrm{Hz}$. The lowest $P_{\mathrm{red}}$ was achieved by \textbf{Proposed} at most frequencies, especially below $400~\mathrm{Hz}$.


\begin{figure}[t]
  \centering
  \centerline{\includegraphics[width=0.9\columnwidth]{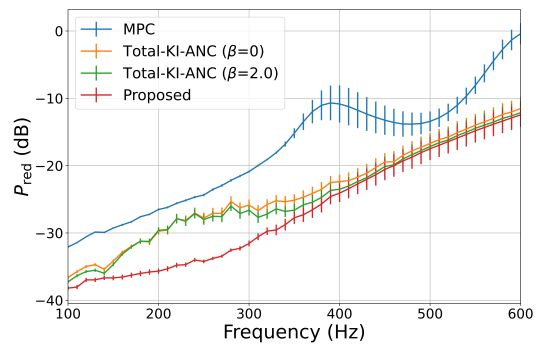}}
\caption{Regional noise power reduction after 12000 iterations with respect to frequency when the primary noise source direction was perturbed.}
\label{fig:4-03}
\end{figure}

To investigate the robustness against the mismatch between the true and prior directions of the primary noise source, we added perturbation to the primary noise source position by using a Gaussian distribution of mean $0$ and standard deviations $0.05~\mathrm{m}$, $6.0~\mathrm{deg}$, and $3.0~\mathrm{deg}$ for its radial, azimuth, and zenith coordinates, respectively. We performed $50$ trials. Fig.~\ref{fig:4-03} shows the mean of $P_{\mathrm{red}}$ after $12000$ iterations with the standard deviation shown as error bars. Although the standard deviation increased for high frequencies, the standard deviation values of \textbf{Proposed} and \textbf{Total-KI-ANC} for $\beta=0.0$ and $2.0$ were almost the same. 
Although \textbf{MPC} does not explicitly consider the direction of the primary noise source, its performance also fluctuates as largely as the other methods. It can be considered that this fluctuation originates from the change of the primary sound field inside the target region.
Even when the prior direction included errors, the mean of $P_{\mathrm{red}}$ of \textbf{Proposed} was still the lowest among the four methods at most frequencies. 
\vspace{-0.4mm}
\section{Conclusion}\label{sec:conclusion}
We proposed a spatial ANC method based on the individual kernel interpolation of sound fields. The kernel-interpolation-based spatial ANC method can reduce incoming noise over a spatial target region with flexible array configurations. The kernel interpolation with directional weighting makes it possible to incorporate prior information on the primary noise source directions in the estimation. However, simply applying this interpolation procedure to spatial ANC is inappropriate because the total sound field is estimated by using a single directional weight. The proposed individual kernel interpolation method separately estimates the primary and secondary sound fields, and we formulated an NLMS algorithm based on this interpolation method. In numerical experiments, a large regional noise reduction was achieved by the proposed method compared with the total-kernel-interpolation-based method. 
\section{Acknowledgment}
This work was supported by JST PRESTO Grant Number JPMJPR18J4 and JSPS KAKENHI Grant Number JP19H01116.

\newpage

\bibliographystyle{IEEEbib_mod}
\bibliography{str_def_abrv, koyama_en, reference}
\end{document}